\documentclass{aip-cp}
\usepackage[numbers]{natbib}
\usepackage{rotating}
\usepackage{graphicx}
\usepackage{mathrsfs}
\usepackage{amssymb}
\usepackage[normalem]{ulem}
\usepackage[dvips]{color}
\usepackage{bm}
\usepackage{longtable}
\usepackage{times}
\usepackage[portuges]{babel}
\renewcommand\sout{\bgroup \color{red} \ULdepth=-.5ex \ULset}

\renewcommand{\v}[1]{\textbf{#1}}
\renewcommand{\rm}[1]{\textrm{#1}}

\begin{document}

% Title portion
\title{Neutron star properties: Constraining the nuclear matter EoS}
\author[aff1]{Constan\c ca Provid\^encia\corref{cor1}}
%\author[aff1]{Bao-An Li\corref{cor1}}
\corresp[cor1]{Corresponding author and speaker: cp@uc.pt}
%\author[aff3]{Lie-Wen Chen}
%\eaddress{bjcai87@gmail.com}; \eaddress{Lwchen@sjtu.edu.cn}
\affil[aff1]{CFisUC, Department of Physics, University of Coimbra,
  3004-516 Coimbra,
  Portugal}
%\affil[aff2]{Department of Physics, Shanghai University, Shanghai 200444, China}
%\affil[aff3]{School of Physics and Astronomy and Shanghai Key Laboratory for Particle Physics and Cosmology, Shanghai Jiao Tong University, Shanghai 200240, China}
\maketitle

\begin{abstract}
We examine the influence of the density dependence of the symmetry
energy on several properties of neutron stars. In particular, we study the constraints set on the nuclear matter equation of state
by the values of the tidal deformability and neutron star
radius,  using a diverse set of relativistic and non-relativistic
mean field models consistent with bulk properties of finite nuclei and
the observed lower bound on the maximum mass of neutron star. The
tidal deformability and radius show a strong correlation with specific
linear combinations of the isoscalar and isovector nuclear matter
parameters associated with the EoS. Such correlations suggest that a
precise value of the radius or the tidal deformability can put tight bounds on
several EoS parameters, in particular, on the slope of the
incompressibility and the curvature of the symmetry energy. 
We show that the density dependence of
the symmetry energy has a direct influence on the amount of
strangeness inside  cold dense matter and, consequently, on the direct
Urca process and cooling of neutron stars. We explain the low
luminosity of SAX 1808.4-3658 as a result of hyperonic direct Urca
processes. Finally, we discuss the strong influence of the
density dependence of the symmetry energy on
the extension of the crust-core transition region of a magnetized
neutron star. The increase of the crust and its of complexity,  due to
the magnetic field effect, may
have a role on the glitch mechanism or on the magnetic field decay.

\end{abstract}

% Head 1
\section{Symmetry energy and neutron star properties}\label{sec2} 

The density dependence of the symmetry is affecting strongly the
properties of
asymmetric nuclear matter, including nuclear properties and neutron
star properties  as the
neutron skin thickness of the nucleus, the neutron star radius or the
onset of the Urca process inside neutron stars,  see
\cite{Li2014} for a review.
In the following, we will discuss how the symmetry energy dependence
on the density  determines neutron star properties as its radius or
tidal deformability, or affects properties of hyperonic stellar matter \cite{Cavagnoli11,Providencia13,Providencia19}  and of
magnetized nuclear matter \cite{Fang2016,Fang2017} .

The symmetry energy is reasonably well constrained at saturation or
sub-saturation densities,  see  \cite{Tsang12,Lattimer13,Oertel17}, but little is known about its behavior at
high densities. The existence of correlations between the EoS
properties and NS properties dependent on the behavior of the EoS at
large densities may help constraining the high density EoS. As an
example, the measurement of the radius of a 1.4$M_\odot$ NS star or its tidal
deformability with a small uncertainty would certainly impose strict
constraints on the high density EoS of asymmetric nuclear matter. In
particular, the recent detection of gravitational waves from a binary
neutron star merger, the GW170817 event \cite{Abbott2017,Abbott2018}, has boosted a large
discussion on this topic, see for instance \cite{Radice2017,De2018,AngLi2018,Tews2018,Lim2018,Tuhin2018,Fattoyev2018,BALi2019}.

To describe neutron stars, the EoS state of stellar matter
must be known in the whole range of densities of relevance. Hyperons
may occur in the NS because its appearance will lower the energy of
the system. Since some of the hyperonic species have a non-zero
isospin, the symmetry energy will also affect the onset and abundance
of these hyperons \cite{Providencia13}, and as a consequence, other
properties as the onset of the direct Urca process. A second
problem we will discuss in the following is the interplay between the symmetry energy properties
and the hyperon content of the star.

Since the measurement of two solar mass NS, PSR J$1614-2230$
\cite{Demorest10,j1614a} and PSR J$0348+0432$ \cite{Antoniadis13}, it has been argued that
this large mass would exclude hyperons from the interior of NS because
the EoS would become too soft, the ``hyperon puzzle'',
\cite{Vidana10,Demorest10,Chatterjee15}. However, it has been shown
that it is possible that if a hard enough EoS or an adequate
choice of the hyperon-meson couplings is considered, the two solar
mass constraint does not exclude hyperon inside neutron stars
\cite{Bednarek2011,Weissenborn12,Weissenborn13}.

A third problem we will also discuss is how the symmetry energy defines
the crust-core transition region of a magnetized neutron star.  To
understand the core properties of a NS it is essential to know quite
well the behavior of its  crust and this justifies this study. It was
shown in \cite{Fang2016,Fang2017,Fang2017a} that a strongly magnetized
star might have a complex region at the crust-core transition, which
would  extend the non-homogeneous range of the star to larger
densities and, therefore, to deeper radii inside the star. This may have
direct implications on the  fractional moment of inertia of the crust
and  the impurity parameter of the crust \cite{Andersson2012,Chamel2013}, quantities that determine
the behavior of the glitch mechanism or the decay of the magnetic
field \cite{pons13}.

Most of the study will be carried out in
the framework of relativistic mean-field models (RMF) calibrated to well
accepted nuclear properties resulting from neutron observations,
theoretical calculations or laboratory experiments. 
Whenever necessary, the NS
properties will be calculated from unified inner crust - core EoS
since it has been shown that
non-unified EoS may give rise to large uncertainties on the radius of low mass neutron stars, although essentially  not affecting the mass
\cite{Fortin2016}. The hyperon-meson couplings will be constrained by
hypernuclei properties \cite{Fortin17}.

\section{Brief review of the formalism \label{sec3}} 
In the next section  we will work with both Skyrme forces and RMF
models. in the other two sections,  the whole discussion will be
carried out in the framework of RMF models.
Within the RFM models,  we consider two different sets, one including
non-linear mesonic terms in the Lagrangian density, the NL set, and a
second one, without these terms but introducing 
density dependent coupling constants, the DD set.
We start from the following Lagrangian density
$
{\mathcal L} = {\mathcal L}_b + {\mathcal L}_m + {\mathcal L}_{m-nl},$
%\label{lagrangian}
%\end{equation}
where the terms $ {\mathcal L}_b, {\mathcal L}_m, {\mathcal L}_{m-nl}$
describe, respectively, the baryons interacting with the mesons, the free
mesons, and  self-interaction and
non-linear mixing terms involving mesons, these last terms only
present in the NL class of models. The different terms are given by \cite{Providencia19}
\begin{eqnarray}
{\mathcal L}_b &=& \sum_{j=1}^8  \bar \psi_j \left( i \gamma_\mu \partial^\mu - m_j +
  g_{\sigma j} \sigma
  + g_{\sigma^* j} \sigma^*- g_{\omega j} \gamma_\mu \omega^\mu - g_{\phi j}
\gamma_\mu \phi^\mu - g_{\rho j} \gamma_\mu \vec{\rho}^\mu
   \vec{I}_j\right) \psi_j,
\end{eqnarray}
\begin{eqnarray}
{\mathcal L}_m 
&=&
+  \frac{1}{2} (\partial_\mu \sigma \partial^\mu \sigma - m_\sigma^2 \sigma^2) %\nonumber \\
%  - \frac{1}{3} g_2 \sigma^3 - \frac{1}{4} g_3 \sigma^4 \nonumber \\
+  \frac{1}{2} (\partial_\mu \sigma^* \partial^\mu \sigma^* - m_{\sigma^*}^2
  {\sigma^*}^2) %\nonumber \\ &&
%+ \frac{1}{4} c_3 (\omega_\mu \omega^\mu)^2 + {\mathcal L}_{nl}\nonumber \\ && 
- \frac{1}{4}
W_{\mu\nu} W^{\mu\nu} 
- \frac{1}{4}
P_{\mu\nu} P^{\mu\nu} 
- \frac{1}{4}
\vec{R}_{\mu\nu} \vec{R}^{\mu\nu} \nonumber \\ && 
+ \frac{1}{2} m^2_\omega \omega_\mu \omega^\mu 
+ \frac{1}{2} m^2_\phi \phi_\mu \phi^\mu 
+ \frac{1}{2} m^2_\rho \vec{\rho}_\mu \cdot \vec{\rho}^\mu ~,
\end{eqnarray}
\begin{eqnarray}
{\mathcal L}_{m-nl}(\sigma,\omega_\mu \omega^\mu) &= &
 - \frac{1}{3} g_2 \sigma^3 - \frac{1}{4} g_3 \sigma^4 + \frac{1}{4} c_3 (\omega_\mu \omega^\mu)^2 
+\left(a_1 g_{\sigma N}^2
  \sigma^2 + b_1 g_{\omega N}\omega_\mu \omega^\mu\right) \vec{\rho}_\mu
\cdot \vec{\rho}^\mu 
% \nonumber \\ &=&-U_s(\sigma)+U_v(\omega^\mu\omega_\mu)+ A(\sigma,\omega^\mu\omega_\mu) \vec{\rho}_\mu\cdot \vec{\rho}^\mu.
\end{eqnarray}
 In the above expressions, $\psi_j$ describes  baryon $j$, 
$\sigma, \sigma^*$ are scalar-isoscalar  meson
fields, and $\omega^\mu$, $\phi^\mu$, $\vec\rho^\mu$ stand for 
the vector isoscalar and isovector fields, and
 $W_{\mu\nu}, P_{\mu\nu}, \vec{R}_{\mu\nu}$ are the
vector meson field tensors
$
V_{\mu\nu} = \partial_\mu V_\nu - \partial_\nu V_\mu.
$

The couplings $g_{\sigma N}$, $g_{\omega N}$, $g_{\rho  N}$  in DD
models are
functions of the baryonic density fitted to nuclear matter properties, see \cite{typel10,ddme2}. 
For NL models, the 
  coupling constants  $g_{\sigma N}$, $g_{\omega N}$, $g_{\rho
  N}$, $g_2$, $g_3$, $c_3$, $a_1$, $b_1$
and  the $\sigma$, $\omega$ and $\rho$ meson masses are fitted to
 experimental, theoretical and
observational data.

Besides the  RMF models described
by the above Lagrangian density, we will also consider Skyrme forces
in  the next  section, all predicting two solar mass stars. As a
second constraint, we require that these models
are still causal in the center of  stars with a mass equal to
$2M_\odot$. The values of the EoS parameters  of these models at saturation density
vary in a wide range of values and are given in \cite{Fortin2016,Providencia19}.

In the section on the hyperonic EoS, we will consider
in the class of  DD models the models DD2 \cite{typel10}  and
DDME2 \cite{ddme2}. In the class of NL models we choose
FSU2 \cite{Chen2014}, FSU2H and FSU2R \cite{Tolos17,Negreiros18}, NL3
\cite{nl3}, NL3 $\sigma\rho$ and NL3 $\omega\rho$
\cite{Pais16,Horowitz2001}, TM1 \cite{tm1}, TM1$\omega\rho$ and
TM1$\sigma\rho$ \cite{Pais16,Bao2014}, TM1-2 and TM1-2 $\omega\rho$
\cite{Providencia13}. Finally, in the last  section, the models  NL3
and NL3 $\omega\rho$ will be used.

\section{Symmetry energy and neutron star radius and tidal
  deformability \label{sec4}}

   In the present section we discuss the existing  correlations between
   the neutron star radii, the tidal deformability,  the Love number
   and several EoS parameters, or linear combinations of parameters,
   evaluated at saturation:  the incompressibility $K_0$, the skewness
   $Q_0$, the slope of the incompressibility $M_0$, the symmetry
   energy $J_0$, its slope $L_0$, and its curvature
   $K_{sym,0}$. In Ref. \cite{Fortin2016} it was
   shown that non-unified EoSs may
introduce a large uncertainty on the determination of low-mass star
radii, therefore, in the following we will consider full unified EoS
for the  Skyrme EoS  and a 
unified inner crust-core EoS for the RMF models. For these models  we have
taken for the outer crust  the EoS proposed in Ref.\
  \cite{ruester06},  
and the EoS of the inner crust was calculated performing a
  Thomas Fermi calculation using the same
  model of the core EoS, see \cite{Grill12,Grill14}. For more details
  on this point please see  Ref. \cite{Fortin2016,Alam2016}.

\begin{figure}[ht!]
\centering
\begin{tabular}{cc}
    \includegraphics[width=.4\textwidth,angle=0]{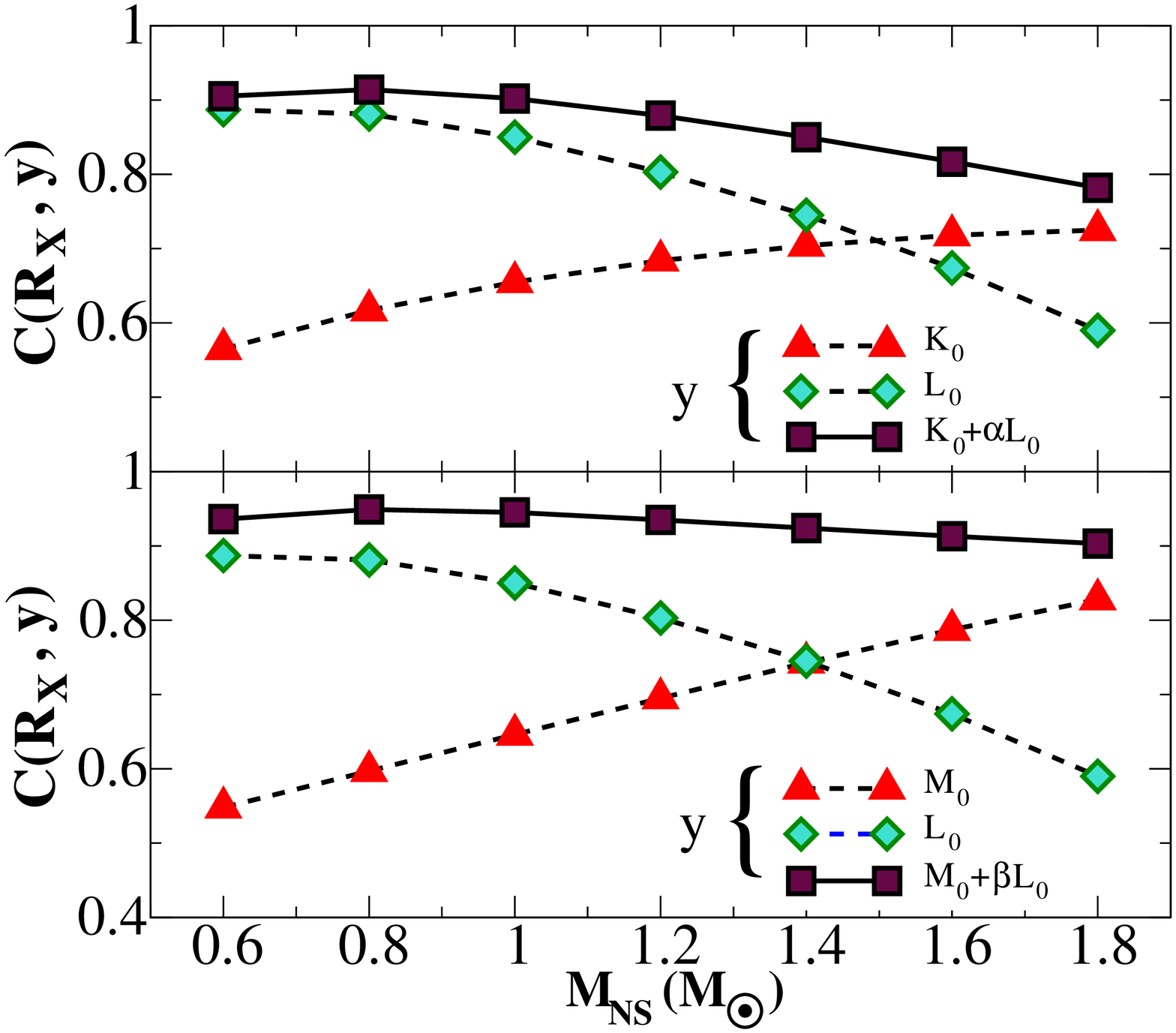}
    \includegraphics[width=.5\textwidth,angle=0]{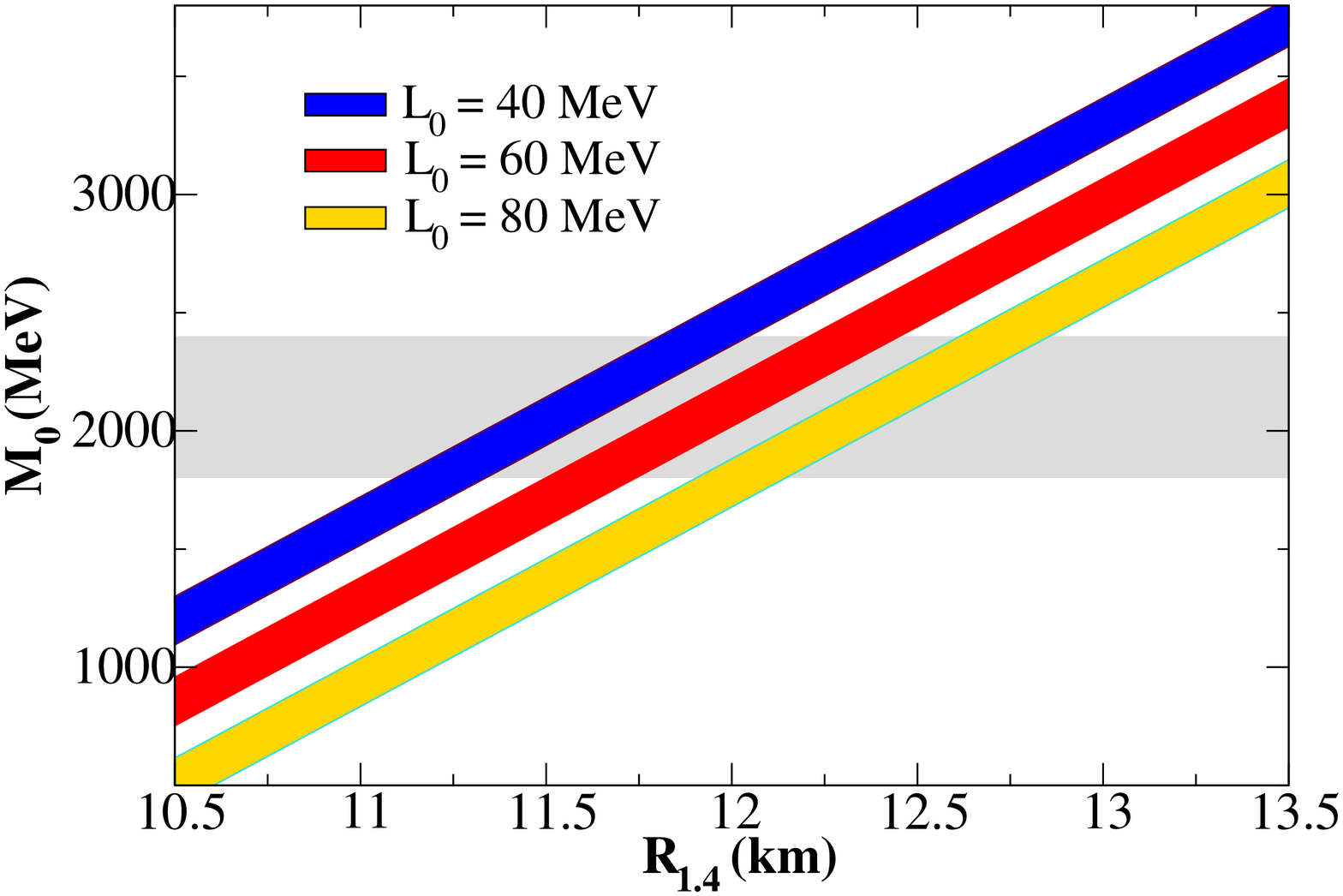}
\end{tabular}
  \caption{ Left panel: The correlation coefficients between the neutron star radii
 and the '$y$' EoS parameters as a function of the neutron star mass,
 where the 
'$y$'  parameters denote $K_0$, $L_0$, and $K_0+\alpha L_0$ 
in the top panel, and $M_0$, $L_0$, and $M_0+\beta L_0$ in the bottom
panel. Right panel:  $M_0$ as a function of $R_{1.4}$ for $L_0=40$, $60$ and $80$ MeV, as obtained
  from the multiple linear regression. The gray shaded region indicates the
   constraint on $M_0$ derived in Ref. \cite{De2015}.
Adapted from \cite{Alam2016} }
  \label{Lradius}
\end{figure}

The linear correlation 
 between any two quantities, $a$ and $b$,  will be calculated by  the Pearson's correlation coefficient, $C(a,b)$, given by
\begin{equation}
C(a,b)=\frac{\sigma_{ab}}{\sqrt{\sigma_{aa}\sigma_{bb}}}\, ,
\quad\mbox{with} % \\
% \label{eq:cc}
% \end{equation}
% with the covariance, $\sigma_{ab}$, written as
% \begin{equation}
\quad \sigma_{ab}=\frac{1}{N_m}\sum_i a_i b_i -\left(\frac{1}{N_m}\sum_i a_i\right
)\left(\frac{1}{N_m}\sum_i b_i\right ) \, .
\end{equation}
In this expression the index $i$ runs over the $N_m$ number of models \cite{Alam2016,Tuhin2018}.
The quantities $a_i$ and $b_i$ correspond,  respectively, to the  neutron star radius for
a fixed mass and an EoS parameter or a linear combination of two parameters.

In Fig. \ref{Lradius} we present,  as a function of the star mass, results for the correlation between
the radius and a) $K_0$, $L_0$ and the linear combination $K_0+\alpha
L_0$, where $\alpha$ is chosen for each mass so as to give the largest
correlation (top panel) and b) $M_0$, $L_0$ and the linear combination $M_0+\eta
L_0$, where $\eta$ and $\alpha$ are chosen in order to maximize the
correlation. The correlation between $L_0$ and the
radius is quite large only for low mass neutron stars. This same result
was confirmed in \cite{Tuhin2018}. For large masses a better correlation
is obtained with the incompressibility $K_0$ at saturation or its
slope $M_0$.  An improved
correlation in the whole range of masses is obtained taking the combination $M_0+\eta
L_0$, or although not so strong, the combination $K_0+\alpha
L_0$. These results indicate that the knowledge of the EoS parameters,
$K_0$, $M_0$ and $L_0$ allows a good estimation of the star radius. 
In the right
panel of  Fig. \ref{Lradius}, we demonstrate how we can obtain the star
radius from  $L_0$ and $M_0$ with the indication of uncertainties,
taking $M_0=1800-2400$ \cite{De2015} and $L_0$ within the range
$40<L_0<80$ MeV, as
suggested by \cite{Tsang12,Lattimer2013} and \cite{Oertel17}. 
In  this figure, 
the incompressibility slope $M_0$  is plotted as a function
of the $1.4M_\odot$ radius, $R_{1.4}$,
for three values of the symmetry energy slope $L_0=40$, $60$ and
$80$ MeV, and the  gray shaded region denotes the constraint on
$M_0$ determined in Ref. \cite{De2015}.  We have obtained
$11.09\lesssim R_{1.4}\lesssim12.86$  consistent with others
predictions as in  Ref. \cite{Li2006,Steiner16}.

\begin{figure}[ht!]
\begin{tabular}{cc}
\includegraphics[width=0.45\linewidth]{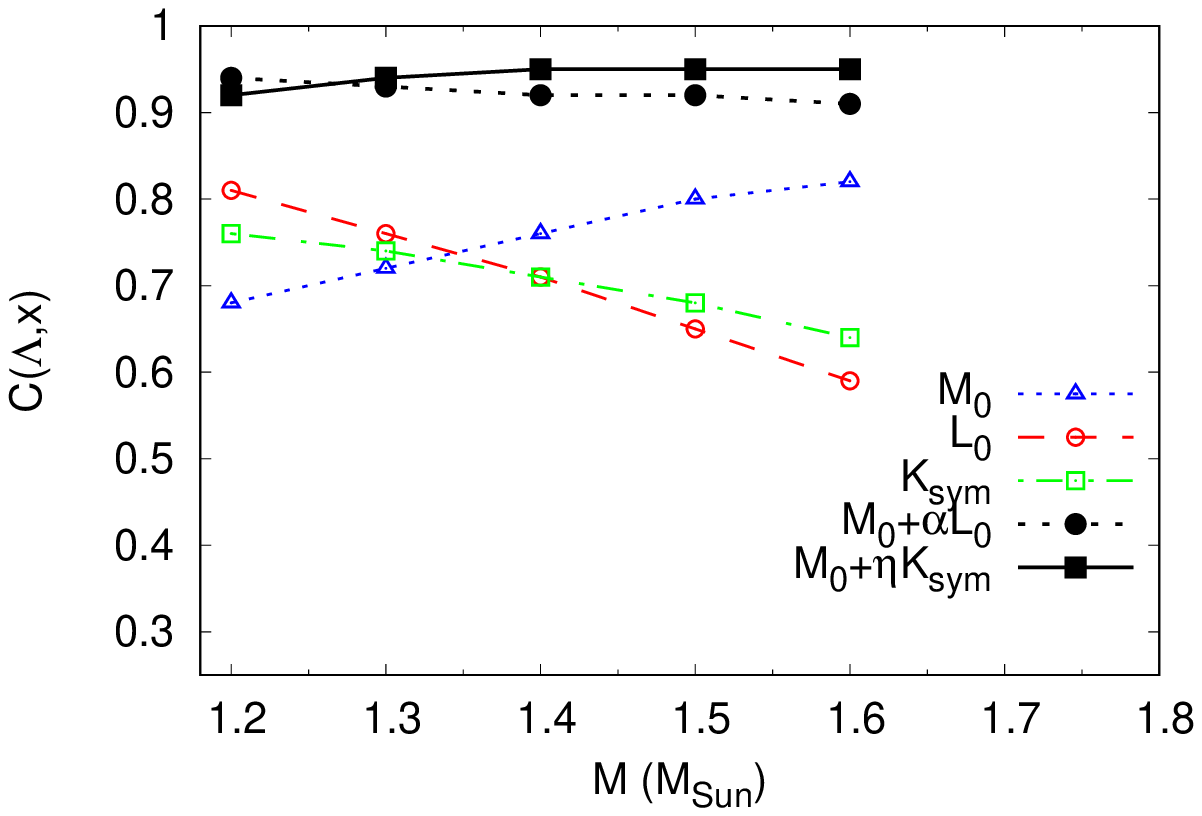}
\includegraphics[width=0.45\linewidth]{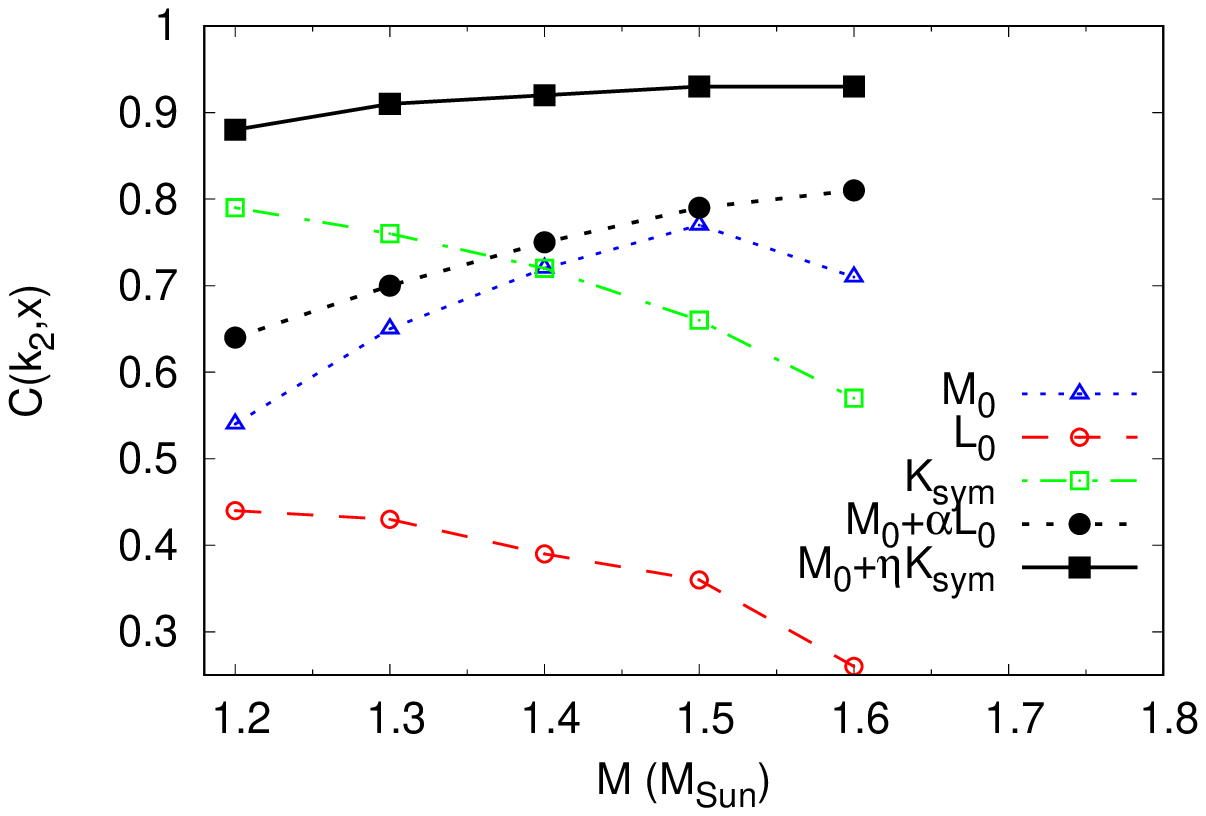}
\end{tabular}
\caption{Pearson correlation coefficients for the correlations between
the tidal deformability (left) and Love number $k_2$ (right), evaluated
for NS with masses between 1.2 and 1.6 $M_\odot$, and the EoS
properties: $M_0$, $L_0$ and $K_{sym}$. Data from \cite{Tuhin2018} }
\label{tidal}
\end{figure}

In \cite{Tuhin2018} the Pearson correlations between the dimensionless tidal
deformability $\Lambda= \frac{2}{3}k_2\left(\frac{R}{M}\right)^5$ and Love
  number  the $k_2$ \cite{Hinderer} of stars with $1.2\le M \le1.6 M_\odot$ have also been calculated for the
  same set of models.  Only  weak or moderate correlations have been obtained between
$\Lambda$, $k_2$ and  the EoS parameters . In Fig. \ref{tidal}, these
correlations are shown for $M_0$, $L_0$ and $K_{sym}$. However,
similarly to the discussion above for the NS radius, also for the
tidal deformability a strong correlation is determined  with the
linear combinations $M_0+\beta L_0$ and $M_0+ \eta K_{\rm sym,0}$ over a 
wide range of NS masses. The parameters $\beta$ and $\eta$ are fixed
so as to maximize the correlation and can be parametrized as $\beta=-1.90 + 265.02~\exp(-M/0.49)$ and $\eta= -1.4 +
29.81~ \exp(-M/0.89)$, where $M$ is the NS mass
in units of solar mass. Using the linear combination $M_0+ \eta K_{\rm
  sym,0}$ together with the upper and lower bounds on
$\Lambda_{1.4}$ set, respectively,  by the GW170817
detection   \cite{Abbott2017,De2018}  and  by the
interpretation of its UV/optical/infrared counterpart, 
together with the empirical ranges of $L_0$ presented in Ref. \cite{Oertel17},
the values of $M_0$ and $K_{\rm sym,0}$ may be constrained 
to the intervals   $1926<M_0<3768$ MeV and $-140<K_{\rm sym,0}<16$ MeV \cite{Tuhin2018}.
.

\section{Symmetry energy and hyperonic neutron stars \label{sec5}}

In the present section, we discuss the effect of the density
dependence of the symmetry
energy on the properties of hyperonic
stars, in particular, on the direct Urca process (DU) both nucleonic
and hyperonic, and on the onset of the different hyperonic species.
The study will be undertaken applying the TM1 model   \cite{tm1} with an extra
non-linear term that couples the $\omega$ and $\rho$-mesons allowing
the change of the density dependence of the symmetry energy. We,
therefore,  generate a family of models with the same 
isoscalar properties but different isovector properties, e.g.  the slope of the
symmetry energy at saturation varies in the range $55<L<110$ MeV
\cite{Bao2014,Pais16}.

The  hyperon-meson couplings have been chosen so that: a) the coupling
to the $\rho$-meson is defined by the isospin projection times
nucleon-$\rho$-meson coupling; b) the couplings to the vector-mesons are fixed considering
the  SU(6) quark model prediction; c) the coupling
$\Lambda$-$\sigma$-meson is fitted to hypernuclei data as in
\cite{Fortin17}, for the $\Sigma$-meson we consider a set of the
different couplings which we will discuss, and for the $\Xi$-meson we
take the $\Xi$-potential in symmetric nuclear matter at saturation
 -18 MeV; d) we include the $\phi$-meson but not the $\sigma^*$
meson except for the $\Lambda$-hyperon.

We first discuss the  nucleonic electron
direct Urca (DU) process \cite{DU91}, the most efficient process when
discussing the cooling on NS. In order to operate, both energy and
linear momentum must be conserved.  The second
conservation law shifts the onset of this process to quite large
densities due to the large neutron  fraction together with a small
proton fraction. In particular, the minimum proton fraction for the
process to occur is $
Y_{p}^{\rm {min}}=\frac{1}{1+\left(1+x_e^{1/3}\right)^{3}}$, with
$x_e=n_e/\left(n_e+n_\mu\right)$,  and $n_e$ and  $n_\mu$   the
electron and muon  densities. In the following, 
$n_{\rm {DU}}$ and mass $M_{\rm DU}$ designate, respectively, the baryonic density
at which the DU process opens and the mass of the star  with a central
density equal to $n_{\rm{DU}}$.

We perform the study within the RMF models presented in Section
\ref{sec3}. These models fall in three different categories: a) models
like TM1, NL3, FSU2, which have a large symmetry energy slope predict, the
opening of the process at densities below the hyperon onset,
corresponding to quite low mass stars; b) models like DD2 and DDME2
completely exclude this process inside the stars; c) models, lke
FSU2R, FSU2H, NL3$\omega\rho$ or TM1$\omega\rho$ with a symmetry
energy slope below 75 MeV,  
predict the onset of the process above the onset of the hyperon onset.

\begin{figure}[ht!]
\centering
\begin{tabular}{cc}
  \includegraphics[width=.5\textwidth,angle=0]{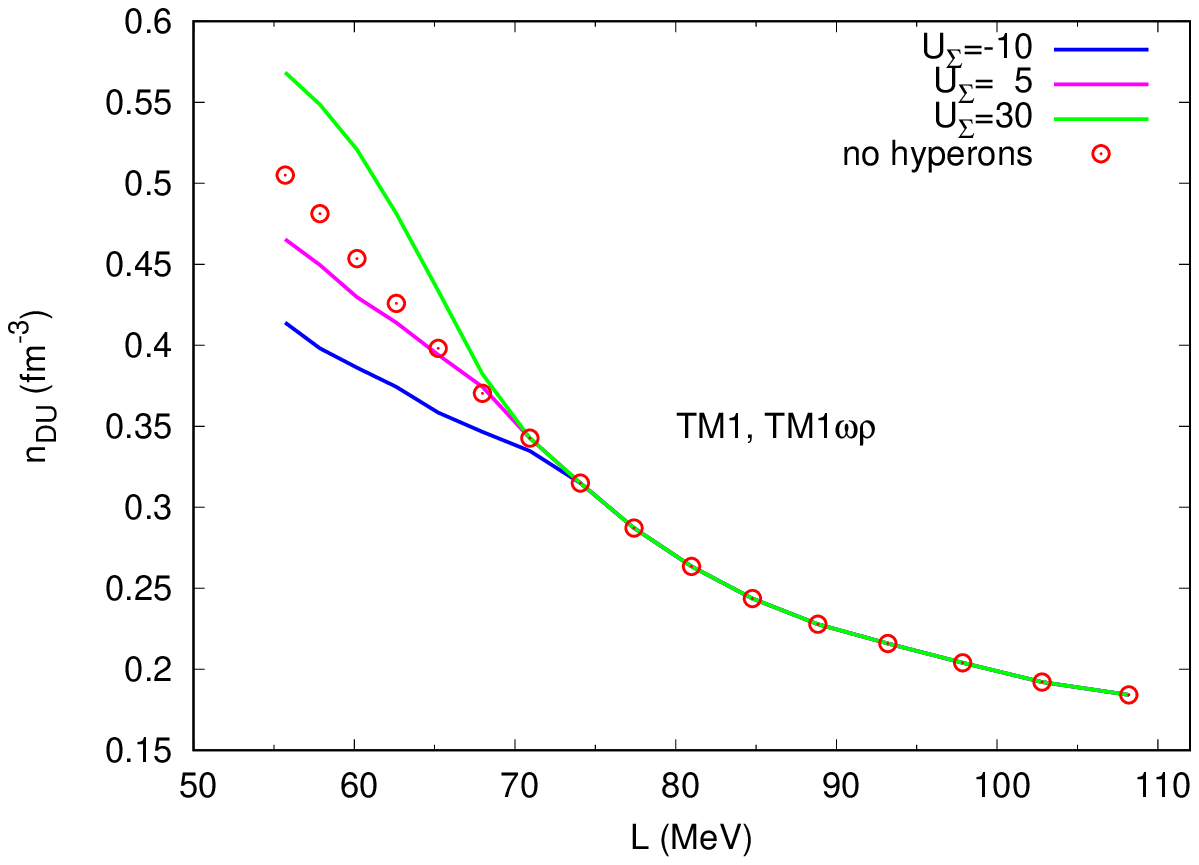}& 
 \hspace{-0.5cm}    \includegraphics[width=.5\textwidth,angle=0]{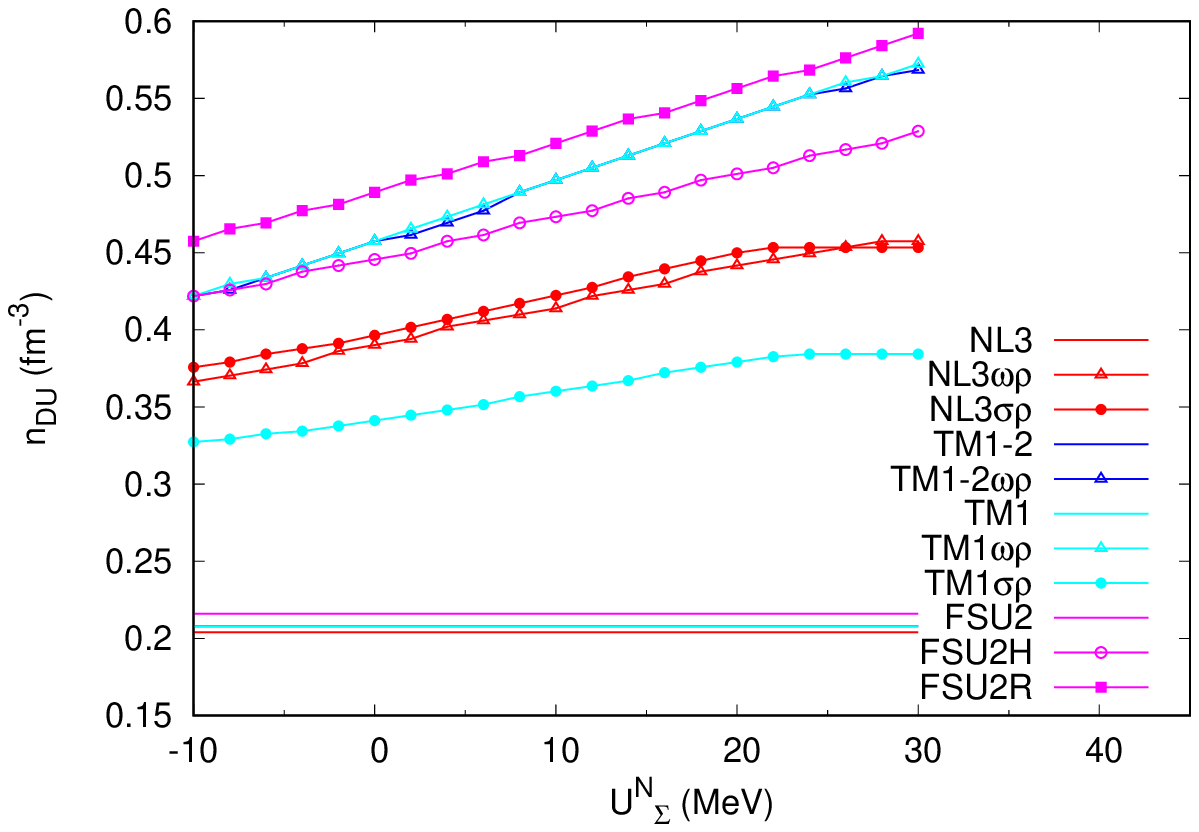}
\end{tabular}
  \caption{Direct Urca process onset density  (left panel) as a function of the symmetry
    energy slope for the TM1$\omega\rho$ models and several values of
    the $\Sigma$
    potential, 
    $U_\Sigma$; (middle panel)  as a function of the $U_\Sigma$ for several models that predict DUrca in the
    core of a NS; (right panel) the onset density of the different
    hyperonic species as a function of the symmetry energy for different values of $U_\Sigma$.}
  \label{durca}
\end{figure}

In Fig. \ref{durca} we plot with open circles the onset
density of the nucleonic electron DU as a function of the slope $L$ for
the family of TM1$\omega\rho$ models.
Since the proton fraction inside the star is defined
by the density dependence of the symmetry energy, it results that
 there is a strong dependence of the DU onset on $L$, as previously discussed
in \cite{Horowitz02,Cavagnoli11}. Models with a small $L$ at
saturation have smaller symmetry energies at supra-saturation
densities, favor larger neutron-proton asymmetries and,
as a consequence, disfavor the onset of DU at small densities.

In the presence of hyperons, 
channels for neutrino emission involving the hyperons are also possible  \cite{Prakash92}. Moreover, the hyperons will
also affect the nucleonic electron DU process, because the fraction of
electrons is strongly dependent on the presence of negatively charged
hyperons. This is shown by the full lines in Fig. \ref{durca}, which
were calculated allowing for
the appearance of hyperons,  each one corresponding to a different
$\Sigma$ potential in symmetric nuclear matter. The effect is only
present for $L\lesssim 75$ MeV: the more repulsive the $\Sigma$
potential, the larger the onset density of the nucleonic DU. For an
attractive or only slightly repulsive potential the onset of the DU
process is smaller than the corresponding density for a pure nucleonic system.
In fact,  the $\Sigma^-$ hyperon is the first to set in and, as a
consequence, the proton fraction increases, the difference between
the proton and neutron Fermi momenta decreases and the DU is  favored \cite{Providencia19}. 
For a more repulsive $\Sigma$ potential, the $\Lambda$-hyperon sets in
first. As soon as it appears neutrons, protons, electrons and muons all suffer a
reduction, and the overall effect is to disfavor the DU
process as compared to the pure nucleonic one.

The most efficient hyperonic DU processes are described by the equations:
\begin{eqnarray}
\Sigma^-&\to&\Sigma^0\ell^- \bar \nu_\ell,\,\quad R=0.61 \label{y2}\\
\Xi^-&\to& \Xi^0 \ell^- \bar \nu_\ell,\,\quad R=0.22 \label{y5}\\
\Sigma^-&\to& \Lambda \ell^- \bar \nu_\ell,\,\quad R=0.21 \label{y3}
\end{eqnarray}
where the $R$ factor indicates the efficiency of the process compared
with  the nucleonic DU process (see \cite{Prakash92}). Two of the
above processes, (\ref{y2}) and (\ref{y3}), involve a $\Sigma$
hyperon, and, in particular, the first is three times more efficient
than the other two. Determining if the $\Sigma$ hyperon is present
inside the NS and which is its abundances, is important to describe the
cooling of a NS. Since this hyperon has isospin one, it is expectable
that its presence is sensitive to the behavior of the symmetry energy.

In the right panel of Fig. \ref{durca}  we plot the onset density of
the nucleonic DU process as a function of the $\Sigma$ potential for
the set of models introduced in Section \ref{sec3} (the DD models do
not predict the nucleonic DU process). For models with a large $L$, as
NL3, TM1 and FSU2, the onset density of DU, $n_{DU}$, does not depend on
$U_\Sigma$ because this density lies below the onset density of
hyperons. For all the other models the more repulsive $U_\Sigma$ the
larger $n_{DU}$, and, as a consequence, the larger the mass of the NS
where it becomes effective.

\begin{figure}[ht!]
\centering
\begin{tabular}{c}
 \includegraphics[width=.6\textwidth,angle=0]{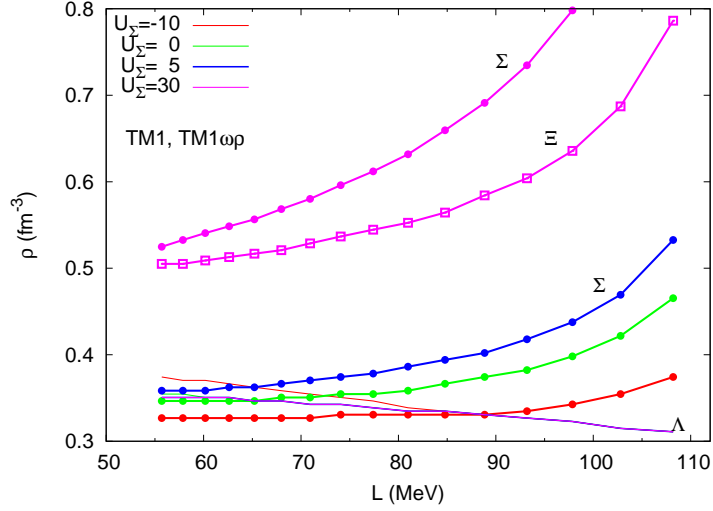}
\end{tabular}
  \caption{ Onset density of the different
    hyperonic species, $\Sigma$ (lines with dots), $\Xi$ (lines with
    open squares), $\Lambda$ (thin lines), as a function of the symmetry energy slope $L$ for
    different values of the $\Sigma$ potential $U_\Sigma$ (given in MeV). The models TM1 and TM1
    $\omega\rho$ have been used in the calculation.}
  \label{hyperons}
\end{figure}

We finally comment on the effect of $L$ on the hyperonic species
inside the star, see  Fig. \ref{hyperons}. The $\Lambda$ hyperon onset is practically not affected by
the value of $U_\Sigma$, and, although its onset density increases
slightly when $L$ decreases, the mass of star at the $\Lambda$-onset
is essentially independent of $L$ and equal to 1.3$M_\odot$. 
The effect of $L$ on the hyperonic species inside the star can be
summarized as follows: a) the  onset of the $\Lambda$ hyperon, which has
zero isospin, is almost not affected by the density dependence of the
symmetry energy; b) the contrary is true for the $\Sigma^-$ and
$\Xi^-$  hyperons which have a non-zero isospin. For these hyperons
their onset density decreases  as $L$ decreases. The $\Sigma$
potential also influences the onset of the other species: the  $\Xi^-$
onset only occurs inside the NS if the $\Sigma$ potential is very
repulsive; the effect is larger for the smaller values of $L$ when the
onset density may be small enough to allow hyperons in stars with
masses only slightly larger than 1$M_\odot$.

\begin{figure}[ht!]
\begin{tabular}{cc}
\includegraphics[width=0.65\linewidth]{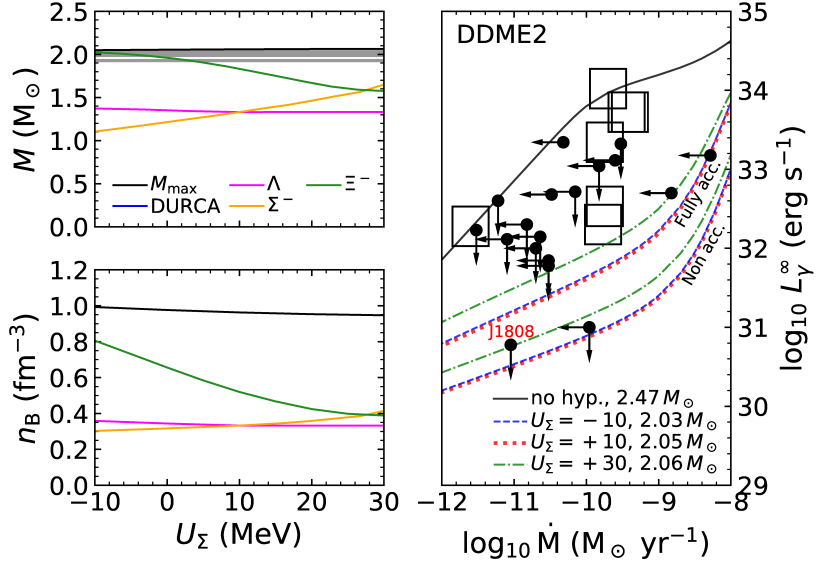}
\end{tabular}
\caption{Left panels: onset densities of the different hyperons and NS
  central baryonic density of maximum mass star (bottom), mass of the
  NS with these densities at the center. Right panel: the luminosity
  of NSs in SXTs obtained for the maximum mass and with the nucleonic
  EoS (black curve) and several hyperonic EoS (colored lines) versus the
  observational data taken from \cite{BY15}. The star indicated in red
  is the 
 SAX J1808.4-3658 and has the lowest observed luminosity. Adapted from \cite{Providencia19}.}
\label{sigma}
\end{figure}

As an application of the above results, in the following we discuss the
effect of the $U_\Sigma$ on the properties of NS described within the
DDME2 model. In  the left panels of Fig. \ref{sigma} we plot in the bottom panel
the
onset densities of the $\Sigma^-$ (yellow), $\Lambda$ (pink) and
$\Xi^-$ (green), as well as the central baryonic density of the
maximum mass star (black) and
in the top panel
the mass of the stars that have in the centre these same densities
as a function of the $\Sigma$ potential. As discussed before, the
onset density for the $\Lambda$ is insensitive to $U_\Sigma$. In this
model the $\Sigma^-$ sets in before the $\Lambda$ for $U\lesssim +10$
MeV, and for $U\lesssim +30$, $\Xi^-$ is the last hyperon to set
in. In any case, the two solar mass constraint is satisfied and
central densities are of the order of 1 fm$^{-3}$ or just below. No
nucleonic electron DU channel will open inside a NS, but the hyperonic
DU process will be active as soon as the different species appear. In
particular, the process described by Eq. (\ref{y3}) will be quite active
above a density of the order of two times saturation density
corresponding to a star with a mass $\sim 1.3 M_\odot$.

In the right panel  of  Fig. \ref{sigma},  we show how the hyperon
presence inside NS affects the cooling of the thermal state of NSs in Soft
X-ray transients (SXT).  In this panel we plot  the luminosity in
quiescence as a function of the accretion rate together with the
observational data taken from \cite{BY15}, for purely nucleonic matter
(black curves) and for different values of the $\Sigma$ potential
(dashed, dotted, and dot-dashed lines).
We are interested in understanding
the behavior of  the SXT with the lowest-observed luminosity, SAX
J1808.4-3658 \cite{SAX,Heinke09}, indicated in red in the right panel.
The upper black curve is obtained with the nucleonic EoS for the
maximum mass star configuration. No DU process occurs and, therefore,  it defines
the lowest possible neutrino
losses and the largest luminosity. The lower bounds were
obtained with the hyperonic EoS,  taking $U_\Sigma=-10,+10, +30$
MeV and considering the maximum mass configuration.  These conditions
define the largest neutrino emissions obtained through DU processes
and, as a consequence, the lowest luminosity.  Both results including a full accreted atmosphere or  non-accreted one
are shown. The low luminosity of
the SAX J1808.4-3658 requires a very efficient neutrino emission
processes operating in the core of the NS. A previous explanation for
this low luminosity state required a quite hard EoS and a set of couplings
of the hyperons to the mesons that favored the appearance of the
complete baryonic octet in the NS core  \cite{Yakovlev04}. In contrast
 the present calculation explains the same state not requiring
the nucleonic DU process, and not allowing for all the hyperon
species, in particular, excluding the very efficient process described
by Eq. (\ref{y2}), if  no or only a small amount of
accreted matter is in
the envelope. Further investigation into this problem will be
undertaken in the future.

\section{Symmetry energy and the crust-core transition of magnetized
  stars \label{sec6}}

In this section, we show how the crust-core transition in a NS is strongly
affected by the magnetic field, and how the magnitude of this effect
depends on the symmetry energy.
For a zero magnetic field, it was shown that there is an anti-correlation
between the transition density and the slope of the symmetry energy
\cite{Vidana09,Xu09,Ducoin10,Ducoin11,Newton2013}. 
Since the magnetic field effects are 
sensitive to the amount of protons, we expect that the influence of
the density dependence of the symmetry energy will reflect itself on
the properties of the neutron star, specially in the range  of small 
proton densities. Large proton densities require more
intense magnetic fields to give rise to non-negligible effects.
Inside a magnetar it is probable that the field intensity will be below $10^{18}$G. 

\begin{figure}[ht!]
\begin{tabular}{cc}
\includegraphics[width=0.45\linewidth]{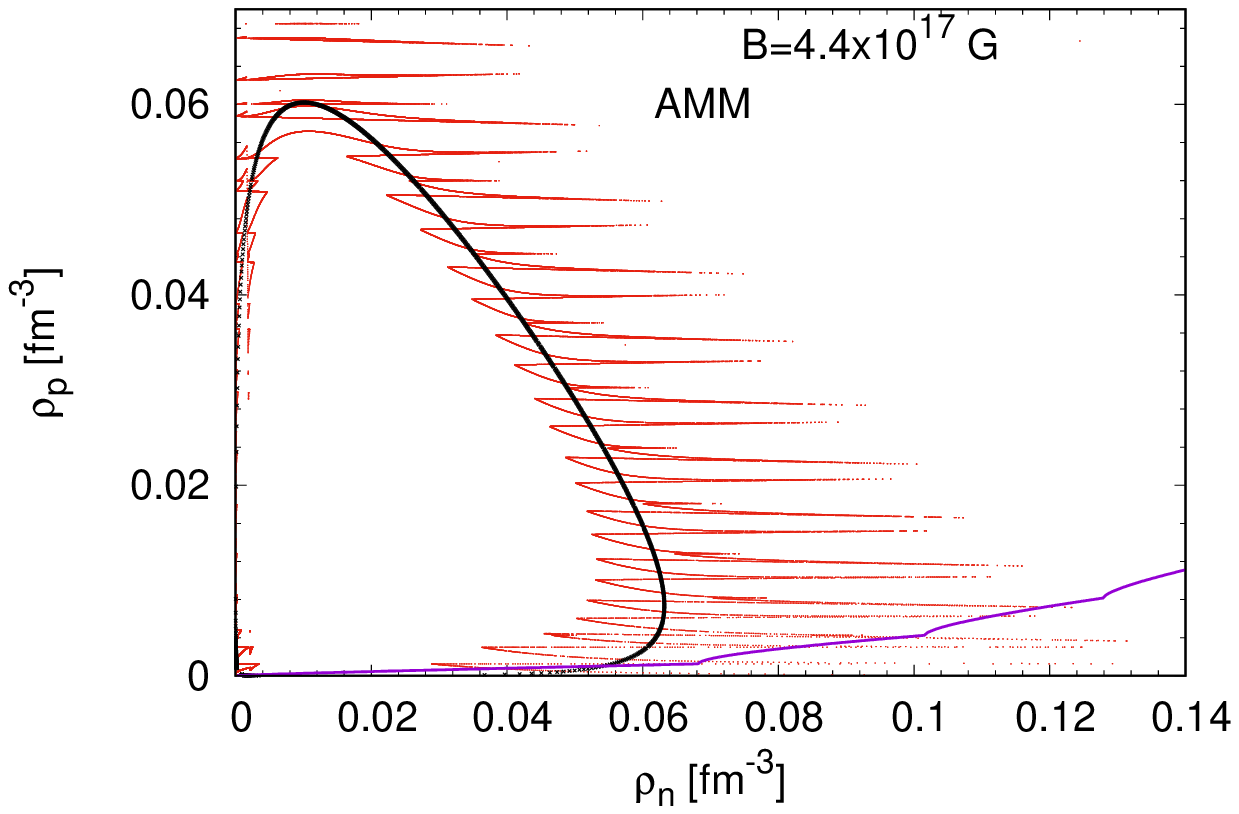} &
\includegraphics[width=0.35\linewidth]{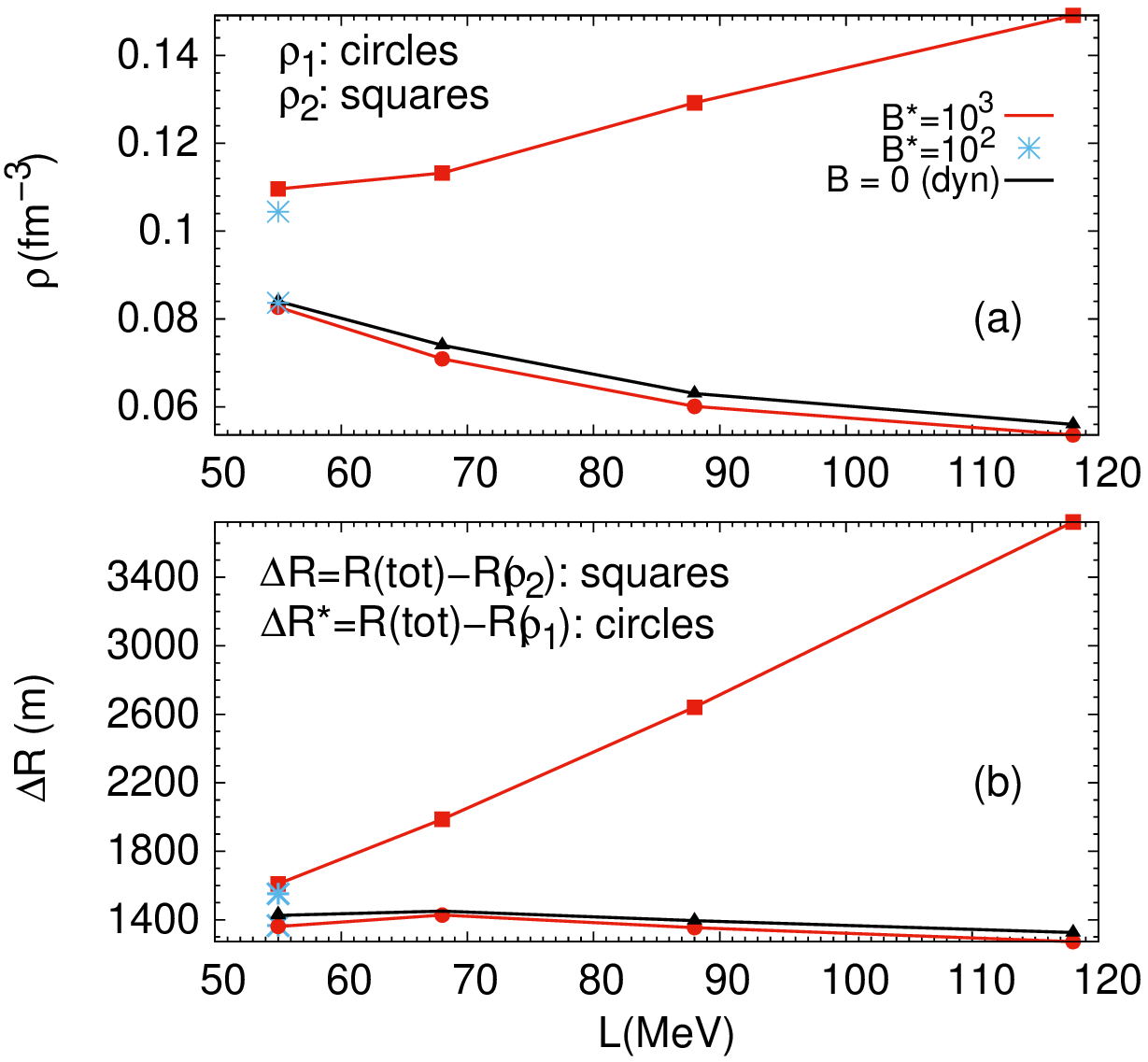} 
\end{tabular}
\caption{Left panel: the dynamical spinodal for nuclear matter obtained
 with $B=0$ (black thick line) and  $B=4.4 \times 10^{17}$G (red
 curve). Also represented is the $\beta$-equilibrium EoS (purple
 line). Right panels:  a) the transition densities, $\rho_1$ and $\rho_2$
 and  b) the crust thickness, $\Delta R$,
  and $\Delta R^*=R({\rm tot})-R(\rho_1)$ 
versus the symmetry energy slope $L$, obtained at $T=0$ with
$B^*=10^3\,\, (B=4.4\times 10^{16}$G)
(red) and $B=0$ (black solid), within the dynamical
 spinodal formalism including the AMM.   For $L=55$ MeV  also $B^*=10^2 (B=4.4\times 10^{15}$)G is shown
 (blue stars). }
\label{spin}
\end{figure}

The  crust-core transition is estimated  from the calculation of the dynamical spinodal
obtained within the relativistic Vlasov equation
formalism applied to relativistic nuclear models
\cite{Providencia2006}. It was shown that this
method gives a good prediction of the crust core-transition in the
absence of a magnetic field \cite{Avancini2010}.
From the analysis of the
dynamical spinodal, we conclude that the crust-core transition does not occur at a
single density, but extends itself into a finite width range of
densities \cite{Fang2016}.  Fig. \ref{spin} illustrates why this
occurs. In this figure the dynamical spinodal section of nuclear
matter at  $B=0$ (black curve) and  for $B=4.4\times 10^{17}$ G  (red
curve) are shown in the $\rho_n-\rho_p$ plane. A complex structure of
bands of 
  unstable matter with large isospin asymmetry are present 
  due to the filling of the Landau levels.  In the same figure, we also
  plot the EoS of $\beta$-equilibrium stellar matter  at zero
  temperature, and it is clearly seen that it crosses the
spinodal section several times. If we associate the range of densities
where these crossings occur to the transition region, we conclude that
this region is formed by a series of alternating stable and unstable 
regions   \cite{Fang2016,Fang2017}.

In the right panel of Fig. \ref{spin}, we   plot the densities
at the bottom and the top of the transition region (red lines in the top panel) and the
radius of the corresponding  neutron star layers as a function of the
slope of the symmetry energy. We have considered a set of models based
on the NL3 parametrization with an extra  non-linear $\omega\rho$
term that introduces a density dependence on the symmetry energy
\cite{Horowitz2001}. The conversion of the density range into  a
neutron star layer was performed in an approximate way,  by  integrating the TOV
equations for non-magnetized stars. In the future this approximation
will be lifted.  In the figures the solid black line corresponds to
the results obtained taking $B=0$. We conclude that the density and the
radius of the bottom layer of the transition density is not much
affected by the magnetic field, and the anti-correlation between the
transition density and the symmetry energy slope is observed.  The top
layer of the transition density, only occurs for a non-zero magnetic
field and its density and radius increases quite fast with the slope
of the symmetry energy. For $L\sim 90$ MeV the transition region as a
thickness of  0.06 fm$^{-3}$ and more than 1 km, reducing to  0.03
fm$^{-3}$ and  400 m if $L=60$ MeV.

The increase of the crust thickness and of the complexity of the crust core transition region may have important
consequences on the properties of neutrons stars, in particular, on
the decay of the magnetic field,  as discussed in \cite{pons13}  or on
the glitch mechanism  \cite{Andersson2012,Chamel2013}.

\section{Conclusions}\label{sec7} 
In the present review,  we have shown how neutron star properties may
set constraints on the EoS of asymmetric nuclear matter at both
sub-saturation  and high densities. We have analysed three different
problems: the correlation between the NS properties, radius, tidal
deformability and Love number, and several EoS parameters or linear
combinations of parameters; the effect of the symmetry energy on the
hyperonic content of neutron stars and on the DU process inside these
stars; and the effect  on the crust-core transition
of magnetized neutron matter.

We could show  that
although the single parameters are generally, at most moderately correlated with
those NS properties, some
well chosen combinations of parameters are very well correlated. This
linear combination together with other constraints either experimental
or observational have allowed to set constraints on  quantities as
the slope of the incompressibility, $M_0$ and the curvature of the
symmetry energy, $K_{sym}$ \cite{Alam2016,Tuhin2018}.

Concerning hyperonic stars, we have shown how the symmetry energy
affects the nucleonic and the hyperonic  direct Urca process.  We have
considered whenever possible constrained hyperon-meson couplings within
a RMF framework \cite{Fortin17}. Due to the lack of data on $\Sigma$-hypernuclei, we
have also analysed how the hyperonic star properties depend on the
$\Sigma$-potential. We have discussed the possibility of explaining
the SXT with the lowest observed luminosity, SAX J1808, as an
hyperonic star described by the DDME2 EoS \cite{Providencia19},  a RMF model with density
dependent couplings \ref{sigma}.

A third problem addressed was the  dependence of the crust-core
transition in strongly magnetized matter on the density dependence of the symmetry energy \cite{Fang2016,Fang2017,Fang2017a}. The
discussion was performed by applying the dynamic spinodal
formalism. In the future, pasta calculations including the magnetic
field effects need to be performed.

\section{Acknowledgement}
I would like to thank Morgane Fortin for preparing the Fig. \ref{sigma}.
This  work  was  supported  by  Funda\c c\~ao  para  a
Ci\^encia e Tecnologia,  Portugal,  under the projects
UID/FIS/04564/2019  and   POCI-01-0145-FEDER-
029912 with financial support from POCI, in its FEDER
component, and by the FCT/MCTES budget through
national  funds  (OE).  Partial  support  comes  also from the
PHAROS COST Action CA16214. 
% References

%  \bibliography{biblio}{}

\begin{thebibliography}{10}

\bibitem{Li2014}
B.-A. Li, A.~Ramos, G.~Verde, and I.~Vidana,
{\em Eur. Phys. J.}, ~A50, 9  (2014).

\bibitem{Cavagnoli11}
R.~Cavagnoli, D.~P. Menezes, and C.~Providencia, 
{\em Phys. Rev.}, ~C84, 065810 (2011).

\bibitem{Providencia13}
C.~Providencia and A.~Rabhi, 
{\em Phys. Rev.}, ~C87,
  055801 (2013).

\bibitem{Providencia19}
C.~Providência, M.~Fortin, H.~Pais, and A.~Rabhi, {\em Frontiers in Astronomy and Space Sciences},
  ~6, 13 (2019).

\bibitem{Fang2016}
J.~Fang, S.~Avancini, H.~Pais, and C.~Providência, 
  {\em Phys. Rev.}, ~C94, 062801 (2016).

\bibitem{Fang2017}
J.~Fang, H.~Pais, S.~Pratapsi, S.~Avancini, J.~Li, and C.~Providência, {\em Phys. Rev.}, ~C95,  045802
  (2017).

\bibitem{Tsang12}
M.~B. Tsang {\em et~al.}, {\em Phys. Rev.}, ~C86, 015803
  (2012).

\bibitem{Lattimer13}
J.~M. Lattimer and Y.~Lim, {\em Astrophys. J.}, ~771, 51 (2013).

\bibitem{Oertel17}
M.~Oertel, M.~Hempel, T.~Klähn, and S.~Typel, {\em Rev. Mod. Phys.}, ~89, 
  015007 (2017).

\bibitem{Abbott2017}
B.~P. Abbott {\em et~al.}, {\em Phys. Rev. Lett.}, ~119,
  161101 (2017).

\bibitem{Abbott2018}
B.~P. Abbott {\em et~al.}, {\em Phys. Rev. Lett.}, ~121,  161101
  (2018).

\bibitem{Radice2017}
D.~Radice, A.~Perego, F.~Zappa, and S.~Bernuzzi, 
  {\em Astrophys. J.}, ~852, L29 (2018).

\bibitem{De2018}
S.~De, D.~Finstad, J.~M. Lattimer, D.~A. Brown, E.~Berger, and C.~M. Biwer,
 {\em Phys. Rev. Lett.}, ~121,  091102 (2018).
\newblock [Erratum: Phys. Rev. Lett.121, 259902(2018)].

\bibitem{AngLi2018}
Z.-Y. Zhu, E.-P. Zhou, and A.~Li, {\em Astrophys. J.}, ~862, 
  98  (2018).

\bibitem{Tews2018}
I.~Tews, J.~Margueron, and S.~Reddy, {\em Phys).
  Rev.}, ~C98, 045804 (2018).

\bibitem{Lim2018}
Y.~Lim and J.~W. Holt, {\em Phys. Rev. Lett.}, ~121, 
  062701 (2018).

\bibitem{Tuhin2018}
T.~Malik, N.~Alam, M.~Fortin, C.~Providência, B.~K. Agrawal, T.~K. Jha,
  B.~Kumar, and S.~K. Patra, {\em Phys).
  Rev.}, ~C98, 035804  (2018).

\bibitem{Fattoyev2018}
F.~J. Fattoyev, J.~Piekarewicz, and C.~J. Horowitz, {\em Phys. Rev. Lett.}, ~120,
 172702 (2018).

\bibitem{BALi2019}
N.-B. Zhang and B.-A. Li, {\em
  Eur. Phys. J.}, ~A55, 39 (2019).

\bibitem{Demorest10}
P.~Demorest, T.~Pennucci, S.~Ransom, M.~Roberts, and J.~Hessels, {\em Nature},
  ~467, 1081 (2010).

\bibitem{j1614a}
Z.~Arzoumanian {\em et~al.}, {\em Astrophys. J. Suppl.}, ~235,
  37 (2018).

\bibitem{Antoniadis13}
J.~Antoniadis {\em et~al.}, {\em Science}, ~340, 6131 (2013).

\bibitem{Vidana10}
I.~Vidana, D.~Logoteta, C.~Providencia, A.~Polls, and I.~Bombaci, {\em EPL}, ~94, 11002 (2011).

\bibitem{Chatterjee15}
D.~Chatterjee and I.~Vidaña, {\em Eur. Phys. J.}, ~A52, 29 (2016).

\bibitem{Bednarek2011}
I.~Bednarek, Haensel, J.~L. Zdunik, M.~Bejger, and R.~Manka, {\em Astron. Astrophys.},
  ~543, A157 (2012).

\bibitem{Weissenborn12}
S.~Weissenborn, D.~Chatterjee, and J.~Schaffner-Bielich, {\em Phys).
  Rev.}, ~C85,  065802 (2012);
\newblock [Erratum: Phys. Rev.C90, 019904 (2014)].

\bibitem{Weissenborn13}
S.~Weissenborn, D.~Chatterjee, and J.~Schaffner-Bielich, {\em Nucl.
  Phys.}, ~A914, 421 (2013).

\bibitem{Fang2017a}
J.~Fang, H.~Pais, S.~Pratapsi, and C.~Providência, {\em Phys. Rev.}, ~C95, 062801 (2017).

\bibitem{Andersson2012}
N.~Andersson, K.~Glampedakis, W.~C.~G. Ho, and C.~M. Espinoza, {\em Phys. Rev. Lett.}, ~109,
  241103 (2012).

\bibitem{Chamel2013}
N.~Chamel, {\em Phys. Rev.
  Lett.}, ~110,  011101 (2013).

\bibitem{pons13}
J.~A. Pons, D.~Vigano', and N.~Rea, {\em Nature Phys.},
  ~9, 431(2013).

\bibitem{Fortin2016}
M.~Fortin, C.~Providencia, A.~R. Raduta, F.~Gulminelli, J.~L. Zdunik,
  Haensel, and M.~Bejger, {\em Phys. Rev.}, ~C94, 
  035804 (2016).

\bibitem{Fortin17}
M.~Fortin, S.~S. Avancini, C.~Providência, and I.~Vidaña, {\em
  Phys. Rev. C}, ~95, 065803 (2017).

\bibitem{typel10}
S.~Typel, G.~R{\"o}pke, T.~Kl{\"a}hn, D.~Blaschke, and H.~Wolter,
    {\em Phys.Rev.}, ~C81, 015803 (2010).

\bibitem{ddme2}
G.~A. Lalazissis, T.~Nik\ifmmode \check{s}\else
  \v{s}\fi{}i\ifmmode~\acute{c}\else \'{c}\fi{}, D.~Vretenar, and Ring,
 {\em Phys. Rev. C}, ~71, 024312 (2005).

\bibitem{Chen2014}
W.-C. Chen and J.~Piekarewicz, {\em Phys. Rev.}, ~C90, 
  044305 (2014).

\bibitem{Tolos17}
L.~Tolos, M.~Centelles, and A.~Ramos, {\em Publ. Astron. Soc.
  Austral.}, ~34, e065 (2017).

\bibitem{Negreiros18}
R.~Negreiros, L.~Tolos, M.~Centelles, A.~Ramos, and V.~Dexheimer, {\em Astrophys. J.}, ~863, 
  104 (2018).

\bibitem{nl3}
G.~A. Lalazissis, J.~Konig, and Ring, {\em Phys. Rev.},
  ~C55, 540 (1997).

\bibitem{Pais16}
H.~Pais and C.~Providência, {\em Phys. Rev.}, ~C94, 015808 (2016).

\bibitem{Horowitz2001}
C.~J. Horowitz and J.~Piekarewicz, {\em Phys. Rev. Lett.}, ~86, 5647 (2001).

\bibitem{tm1}
Y.~Sugahara and H.~Toki, {\em Nucl. Phys.}, ~A579,
  557(1994).

\bibitem{Bao2014}
S.~S. Bao and H.~Shen, {\em Phys. Rev.}, ~C89, 
  045807  (2014).

\bibitem{ruester06}
S.~B. Ruester, M.~Hempel, and J.~Schaffner-Bielich, {\em Phys. Rev.}, ~C73, 035804,
  (2006).

\bibitem{Grill12}
F.~Grill, C.~Providencia, and S.~S. Avancini, {\em Phys. Rev.}, ~C85, 055808 (2012).

\bibitem{Grill14}
F.~Grill, H.~Pais, C.~Providência, I.~Vidaña, and S.~S. Avancini, {\em Phys.
  Rev.}, ~C90, 045803 (2014).

\bibitem{Alam2016}
N.~Alam, B.~K. Agrawal, M.~Fortin, H.~Pais, C.~Providência, A.~R. Raduta, and
  A.~Sulaksono, {\em
  Phys. Rev.}, ~C94, 052801 (2016).

\bibitem{De2015}
J.~N. De, S.~K. Samaddar, and B.~K. Agrawal, {\em Phys. Rev.}, ~C92, 014304 (2015).

\bibitem{Lattimer2013}
J.~M. Lattimer and Y.~Lim, {\em Astrophys. J.}, ~771, 51 (2013).

\bibitem{Li2006}
B.-A. Li and A.~W. Steiner, {\em Phys. Lett.}, ~B642,
  436 (2006).

\bibitem{Steiner16}
A.~W. Steiner, J.~M. Lattimer, and E.~F. Brown, {\em Eur. Phys.
  J.}, ~A52, 18 (2016).

\bibitem{Hinderer}
T.~Hinderer, {\em The Astrophysical
  Journal}, ~677,1216 (2008).

\bibitem{DU91}
J.~M. {Lattimer}, C.~J. {Pethick}, M.~{Prakash}, and {Haensel}, {\em Physical Review Letters}, ~66,
  2701 (1991).

\bibitem{Horowitz02}
C.~J. Horowitz and J.~Piekarewicz, {\em Phys. Rev. C},
  ~66, 055803 (2002).

\bibitem{Prakash92}
M.~Prakash, M.~Prakash, J.~M. Lattimer, and C.~J. Pethick, {\em Astrophys. J.},
  ~390, L77 (1992).

\bibitem{BY15}
M.~V. {Beznogov} and D.~G. {Yakovlev}, {\em Monthly Notices of the Royal Astronomical
  Society}, ~447, 1598 (2015).

\bibitem{SAX}
S.~Campana, L.~Stella, F.~Gastaldello, S.~Mereghetti, M.~Colpi, G.~L. Israel,
  L.~Burderi, T.~D. Salvo, and R.~N. Robba, {\em The Astrophysical
  Journal Letters}, ~575,  L15 (2002).

\bibitem{Heinke09}
C.~O. {Heinke}, G. {Jonker}, R.~{Wijnands}, C.~J. {Deloye}, and R.~E.
  {Taam}, {\em The Astrophysical Journal}, ~691, 1035 (2009).

\bibitem{Yakovlev04}
D.~G. Yakovlev and C.~J. Pethick, {\em Ann. Rev.
  Astron. Astrophys.}, ~42, 169  (2004).

\bibitem{Vidana09}
I.~Vidana, C.~Providencia, A.~Polls, and A.~Rios, {\em Phys. Rev.},
  ~C80, 045806 (2009).

\bibitem{Xu09}
J.~Xu, L.-W. Chen, B.-A. Li, and H.-R. Ma, {\em Astrophys. J.}, ~697, 1549
  (2009).

\bibitem{Ducoin10}
C.~Ducoin, J.~Margueron, and C.~Providencia, {\em EPL},
  ~91,  32001 (2010).

\bibitem{Ducoin11}
C.~Ducoin, J.~Margueron, C.~Providencia, and I.~Vidana, {\em
  Phys. Rev.}, ~C83, 045810 (2011).

\bibitem{Newton2013}
W.~G. Newton, M.~Gearheart, and B.-A. Li, {\em Astrophys. J. Suppl.}, ~204, 9 (2013).

\bibitem{Providencia2006}
C.~Providencia, L.~Brito, S.~S. Avancini, D.~P. Menezes, and Chomaz,
  {\em Phys. Rev.}, ~C73, 025805 (2006).

\bibitem{Avancini2010}
S.~S. Avancini, S.~Chiacchiera, D.~P. Menezes, and C.~Providencia, {\em Phys. Rev.}, ~C82,
  055807 (2010).
\newblock [Erratum: Phys. Rev.C85,059904(2012)].

\end{thebibliography}
% \bibliographystyle{ieeetr}
%  \end{document}

 \end{document}